# Long-Term Forecasts of Military Technologies for a 20–30 Year Horizon: An Empirical Assessment of Accuracy

Alexander Kott, U.S. Army Research Laboratory, Adelphi, MD, USA

Philip Perconti, U.S. Army Research Laboratory, Adelphi, MD, USA

Corresponding author: Alexander Kott, U.S. Army Research Laboratory, 2800 Powder Mill Road, RDRL-D, Adelphi, MD 20783-1197; email: alexkott@yahoo.com

Keywords: Technology Forecast; Prediction Accuracy; Military Technology; Long-range Forecast; Forecast Assessment

## Abstract

During the 1990s, while exploring the impact of the collapse of the Soviet Union on developments in future warfare, a number of authors offered forecasts of military technology appearing by the year 2020. This paper offers a quantitative assessment of the accuracy of this group of forecasts. The overall accuracy — by several measures — was assessed as quite high, thereby pointing to the potential value of such forecasts in managing investments in long-term research and development. Major differences in accuracy, with strong statistical significance, were found between forecasts pertaining primarily to information acquisition and processing technologies, as opposed to technologies that aim primarily at physical effects. This paper also proposes several recommendations regarding methodological aspects of forecast accuracy assessments. Although the assessments were restricted to information available in open literature, the expert assessors did not find this constraint a significant detriment to the assessment process.

## Introduction and Motivation

Technology forecasting is increasingly recognized for its value in both commercial and government endeavors. Worldwide, at least 23 organizations perform technology forecasting (Lerner et al. 2015) as their major product for external customers. In addition, virtually every organization engages in some form of forecasting for internal purposes, implicitly or explicitly, in order to plan its activities.

Product developers and manufacturers need technology forecasts in order to know where to invest their product development resources or how to plan new manufacturing facilities. Providers of services, e.g., health care (Doos et al. 2016), use technology forecasts to determine what new equipment should be purchased or whether a purchase should be delayed because the next generation of technology is about to emerge. Governments use technology forecasts to optimize allocation of funds toward supporting educational and scientific research institutions.

This paper concerns itself with a particular type of technology forecast: long-term forecasts of military technology. The reasons such forecasts are valuable to a military establishment are not dissimilar to those that make forecasts important to businesses: a military procurement institution uses technology forecasts to determine what system development and procurement efforts should be undertaken (and funded) to optimize the future value of the resulting technology to those who would have to fight in a future war. If the forecast is wrong, the military may end up fighting an enemy who possesses weapons of superior technology (Van Creveld 2010), not unlike a business that relies on a technology forecast to avoid falling behind its competitors.

On the other hand, in spite of these fundamental similarities, technology forecasting models in business and in certain types of military system do differ by their time horizons. Here, the time horizon is the number of years between the date a technology is predicted to emerge and the date the forecast was made. While most technology forecasts (performed for business purposes) are short term (the horizon is 1–5 years), the technology forecasts for the military in many cases (although there are many important exceptions) is either mid-term (6–10 years) or long term (11–30 years). The necessity of long-term forecasts in the world of military technology is dictated by long periods of time required for full development of some types of complex military systems.

It has become common – and a concern for policy-makers -- for a major defense acquisition program to take on the order of 2 decades (Goure 2017) from concept development to initial operating capability. In addition, development of foundational science and technological knowledge underpinning engineering developments of advanced systems often takes another 10 or more years. In this way, 20 years or even longer horizon is often important for military technology forecasts. This paper, therefore, focuses specifically on long-term (20 years or longer) military technology forecasts.

However, can such long-term forecasts be sufficiently accurate? It is often observed that the longer the horizon of a forecast, the lower the forecast accuracy, e.g., dropping from 38% accuracy for less than 10 years horizon to merely 14% accuracy for horizons over 11 years (Fye et al. 2013). After all, if accuracy is low, there is no value in developing forecasts. Moreover, an inaccurate — and potentially misleading — forecast is worse than no forecast at all. Unfortunately, relevant empirical studies are rare. As we discuss in the next section of this paper, studies of forecast accuracy in general are few. Of these, only a small fraction deals with technology forecasts. Furthermore, virtually all of them deal with short-term, not long-term, forecasts.

The contributions of this study are as follows. First, it adds a valuable empirical data point to a small set of data on long-term forecast accuracy. Second, it constitutes a case study on methodologies suitable for assessing the accuracy of forecasts and yields recommendations for methodological approaches we found to be effective. Finally, it provides support to the argument that long-term technology forecasting is both feasible and sufficiently accurate for the purposes of supporting management decision making on long-term research investments.

## Prior Work

Studies of forecast accuracy exist but are hardly numerous (Doos et al. 2016). For example, Fye et al. (2013) review available studies of forecast accuracy since as early as 1972. These include Armstrong and Grohman (1972), Lorek et al. (1976), Ascher (1979), Makridakis et al. (1979), Carbone et al. (1983), Grove et al. (2000), Makridakis and Hibon (2000), Albright (2002), Armstrong (2006), Berg et al. (2008), and Goldstein and Gigerenzer (2009). Only a few of those studies — e.g., Lerner et al. (2015), Fye et al. (2013), Albright (2002), and Ascher (1979) — are about technologies; others study forecasts of market sizes, corporate earnings, medical outcomes, election outcomes, etc.

Among the technology forecasts, an overwhelming majority focus on a horizon as short as 3–5 years; long-term forecast accuracy assessments are almost non-existent (Lerner et al. 2015). In fact, forecasts with a horizon greater than 20 years are so uncommon that in at least one study they were considered outliers and have been removed from consideration (Fye et al. 2013). In general, longitudinal forecast data are rare. The studies by Albright (2002) and by Parente and Anderson-Parente (2011) are perhaps the only publications with data pertaining to accuracy for long-term (over 25 years) horizons (Rowe and Wright 2011; Halal 2013).

Rowe and Wright (2011) stress the challenge of assessing accuracy of forecasts in general and argue that perhaps the accuracy of a forecast is not truly a good benchmark for forecasting performance; perhaps the consensus reached during the production of a forecast is a more valuable product of the forecasting process. Firat et al. (2008) also report the view that unlike short-term horizons, if the horizon is 15 or more years, the accuracy assessment becomes difficult — and its utility more problematic.

With this, let us consider several examples where the accuracy of forecasts was assessed at a relatively long horizon.

Fye and coworkers (2013) found that technology forecasts for near term (1–5 years horizon) and midterm (6–10 years) were about 38–39% successful, and long term (11 years and longer) only 14% successful. In their study, each forecast stated the year when a particular technology event was predicted to occur. A forecast was considered successful if the forecasted event was realized within ±30% of the total forecasted time horizon around the forecasted year.

Another publication (Halal 2013) reports results of multiple annual validation studies showing that the average variance of all forecasts is +3/-1 years at 10 year horizons. In this case, each forecast asserts the year when a particular technology would reach a 30% adoption level in industrialized countries. The assessment apparently used expert opinions to determine whether the technology indeed reached that adoption level at the horizon time.

A rare study of forecast accuracy at the 30 year horizon (Parente and Anderson-Parente 2011) describes a research when in 1981, 18 scenarios of developments in mental health care were postulated, and 9 of them were predicted to occur while the other 9 were predicted to not occur. In 2011, 5 of these predictions were assessed to be wrong, indicating 72% (13 out of 18) accuracy.

A unique study at the 33 year horizon (Albright 2002) reviews technology forecasts made in 1967 for the year 2000 and finds that forecasts in computers and communication stood out as about 80% correct, while forecasts in all other fields were judged to be less than about 50% correct. We should also note a celebrated example of a 50 year horizon forecast made in 1964 by Isaac Asimov, widely characterized as highly accurate by the press in 2014 (Collins 2014). However, the press' assessments were anecdotal, and we are not aware of a quantitative characterization of those predictions.

## Data: A Set of Long-Term Forecasts

In the 1990s, a number of publications appeared in which authors offered visions of the future of military technologies. Some of them established the year 2020 as their horizon. We conducted rigorous literature searches for works published in 1990–2000 regarding forecasts of warfare and military technology in the year of 2020. These publications ranged broadly from works of fiction to popular articles to organizational internal technical reports and briefs.

From these publications, we excluded those that provided too few specific technical details or where the time horizon was vaguely defined or different from 2020. For example, although Krepinevich's *The Military Technical Revolution: a Preliminary Assessment* (1992) offered a number of insightful and prescient technological military predictions, it did not specify the time horizon. Therefore, we did not include this publication in our research. On the other hand, Hughes' "Future Conditions: The Character and Conduct of War, 2010 and 2020" (2003) targeted the specific 2020 horizon but worded its predictions in exceedingly broad terms that would be difficult to assess in 2020. Other examples, such as Gordon and Wilson (1998), Applegate (2001), Biddle (1998), etc., offered important ideas about the nature and conduct of warfare in 2020 but little in the way of specific technology predictions.

This process left us with the following publications, appropriately diverse in their styles and approaches:

- Newman's (1996) article is a compilation of technology forecasts from a broad range of open sources, intended for a broad audience of mainstream media readers.
- Vickers (1996) is a formal report produced for the Center of Strategic and Budgetary Assessments, which works mainly in support of the US Department of Defense.
- O'Hanlon's (2000) book explores the future technologies of warfare using the author's own analysis (performed in the late 1990s) of scientific foundations of the technologies.
- Scales (1997, 2017) and Scales and Parmentola (1998) describe the findings of the project called an Army After Next performed in 1995–1997 by a group of U.S. Army officers; although not specifically technology-focused, it includes several forecasts of future technologies. Some of these forecasts refer to the year 2025, not 2020.
- Peters (1991) is a work of fiction, a military adventure novel with numerous descriptions of future military technologies. The uncanny accuracy of some of the technologies envisioned in the novel might have stemmed from the author's position as a U.S. Army intelligence officer stationed in Europe at the time of writing the novel.

From these five very different sources we extracted numerous descriptions of anticipated future technologies. In the process of extraction, we excluded technologies related to nuclear, chemical, and biological warfare, as well as discussions of strategic issues. Because the expertise of this paper's authors, and of their expert advisors, relates mainly to matters of ground warfare, we also excluded references to technologies pertaining strictly to naval or air operations.  Furthermore, we excluded statements that were too vague for a meaningful assessment, e.g., "growing role of information". Finally, we excluded statements that referred to such sensitive technologies that would be impossible to evaluate with information only from open literature.

In many cases, forecasts were associated with qualifiers and caveats, such as "it is possible, if sufficient progress is made in research Z, that technology X would become available by 2020." We replaced such statements with a simple assertion that technology X will be available by 2020. In general, the statements were reworded to ensure a degree of consistency, clarity, and brevity.

This process of extraction yielded a total of 81 forecast statements, a few samples of which are shown in Table 1. The entire list is found in the Appendix. The number of statements extracted from each source is shown in Table 2. When similar statements were made by more than one source, we attempted to combine the similar statements into one statement. For this reason, the total number in Table 2 is greater than 81.

*Table 1 Examples of forecast statements. Letters N, V, H, S, and P refer to the source of the forecast (e.g., P for Peters, H for O'Hanlon, etc.).*

| A tank munition will be laser-guided (H) |
|---|
| A robotic (unmanned?) ground vehicle will be used as a scout (P, H) |
| Swarms of armed UAVs (loitering munitions) will be able to destroy numerous ground targets (S, H, V) |

*Table 2 Number of statements included in our analysis, per source.*

| Scales | Peters | Vickers | Newman | O'Hanlon |
|---|---|---|---|---|
| 14 | 12 | 16 | 20 | 33 |

To compare the accuracy of forecasts across different categories of technologies, we grouped the forecast statements into categories that tend to be commonly recognized in the domain of military technologies (see the Appendix where these groups are shown). Defined here very informally, the categories are as follows:
- Line-of-Sight Effects: weapons and other systems that unleash effects at the enemy targets directly visible from the "shooter"
- Non-Line-of-Sight Effects: weapons and other systems that send munitions or other effects at targets that are not directly visible
- Protection: systems used to defeat or mitigate the effects imposed by the enemy on friendly targets
- Platforms: systems that move warriors and systems to and around the battlefield
- Cyber and Electronic Warfare: systems used to interfere with communication, processing, and storage of enemy's information
- Sensing and Information Collection: systems for obtaining information about the enemy and friendly assets
- Command and Control: systems for interpreting the battlefield information, and making and communicating the decisions on actions

In some cases, a forecast statement did not fit neatly into one of the above categories, and we sought the best possible compromise.

In addition to technology forecasts, the sources also offered numerous and generally valuable forecasts on conditions and conduct of military operations in 2020. Many of these forecasts appear to be exceptionally prescient. However, including considerations of such forecasts would take this study well beyond its intended scope.

## Data: Assessing Success of Forecasts

Assessment of a technology forecast's success is typically performed by subject matter experts at the time for which the forecast has been provided, i.e., at or after the horizon time. The experts offer their opinions on whether the technology has had the forecasted degree of impact (e.g., weak, moderate, strong) on a relevant industry at the horizon time (Lerner 2015) or whether the technology has demonstrated an adoption rate forecasted (Fye et al. 2013). The forecast is considered successful if the impact or adoption rate has met or exceeded the forecasted value by the time horizon for which the forecast was made.

Neither approach appeared appropriate for the purposes of our study. In the case of military technologies, it is often impossible to observe a degree of impact or a rate of adoption unless an actual armed conflict of a significant scale breaks out. An extent of fielding is not a useful measure either, because fielding of new military technologies often awaits an actual military necessity. To the extent that we wish to forecast the dynamics of technology per se, as opposed to making a forecast of military-political developments by the horizon time, a different approach is needed.

We elected to focus on the readiness level of the technology (DoD 2009). Specifically, we elected to interpret a forecast that "in 2020, we expect technology X" as a statement that by 2020, technology X will reach the Technology Readiness Level (TRL) of 8, i.e., an actual system has been constructed and qualified through tests and demonstrations. Note that we decided not to require the highest TRL level 9, which calls for proof of the system through successful mission operations, because opportunities to perform mission operations are a function of politico-military circumstances, not of a technology intrinsically. We do not wish to conflate technology dynamics with politico-military dynamics. These two deserve to be forecasted separately. We reason that if a conflict breaks out and a military necessity occurs, a technology of TRL 8 will be available to be pressed into production and operation.

In addition to accepting TRL 8 as a proof-of-existence of a technology, we also consider the technology as "available" if it is offered for sale commercially or if it is successfully fielded at least on an experimental basis. To summarize, in this paper a forecast statement is assessed entirely true if one of the following conditions applies in 2020 or earlier, in any technologically developed country, according to open literature sources:

- The system prototype is demonstrated at TRL 8, or
- The system is offered for sale commercially, or
- The system is in operational use, even if experimental.

Furthermore, we conclude that long-term forecasts do not always lend themselves to binary assessments, i.e., assessing a forecast as either entirely successful or entirely unsuccessful. Because the suitable technology nature and approaches can change drastically over a long-term horizon, for purposes of long-term technology it may be more appropriate to assess a fractional degree to which the actual technology at the horizon time is close to the predicted technology. Consider the following example. Suppose a forecast asserts that in 30 years, technology X will perform functions Y and Z. At the horizon time, 30 years later, the assessor notes that technology X does indeed exhibit the function Y, but due to changes in military environment, function Z is no longer desired and therefore is not available. Should this forecast be considered entirely unsuccessful? Probably not; it should be given at least a "partial credit."

Therefore, in our study we offer experts the following scale of fractional validity of a forecast, both qualitative and quantitative: not true at all (0); somewhat true (0.20–0.25); half true (0.50); mostly true (0.75–0.80); and entirely true (1). In most cases, the experts assessments were binary, i.e., 0 or 1.

It is important to note that the experts were instructed to consider technology developments worldwide. The assessment was not centered on any particular country. A technology was considered available regardless of the country where the development occurred and regardless of which country's military had access to the technology.

A total of 10 experts were engaged in the assessment process. Each expert was a senior scientist-technologist in a position of technical leadership, a civilian employee of the US Army, with 20–30 years of experience in science and technology development and management, and with a doctoral degree. Each expert provided her/his assessments independently and was given the option to assess only those statements that belonged to his/her self-assessed fields of expertise. Therefore, not every forecast was assessed by all 10 assessors. Each forecast statement was assessed by between 4 and 10 assessors, for a total of 637 assessments.

Table 3 shows illustrative excerpts from the results of expert assessments. For each forecast statement, we obtained a minimum of 4 expert assessments, and in some cases up to 9 assessments.

*Table 3 Illustrative excerpts from the results of expert assessments for several of the forecast statements.*

| A tank (or gun, or mortar) will operate unmanned | 1 |   | 1 | 1 | 1 | 0 | .7 | 1 |   | 1 |
|---|---|---|---|---|---|---|---|---|---|---|
| A robotic (unmanned?) ground vehicle will be used as a scout | 1 | 1 | 1 | 1 | 1 | 1 | .7 | 1 |   | .3 |
| A combat air platform will use tilt-rotor | .5 | .5 |   |   | .5 | .5 | .5 | .5 |   | 0 |

While for many forecasts the assessments of all experts were identical, in other cases opinions differed. Rather than simply average the divergent assessments, we performed an adjudication: the prevailing assessments were selected, with special consideration given to the experts who were understood to be most knowledgeable in the field, and to evidence in publications when the evidence was unambiguous. It should be noted that the adjudicated assessments did not necessarily coincide with the opinions of this paper's authors.

Inevitably, the assessments were challenged by the fact that the forecasts were often made in a rather vague manner. Sometimes it was difficult to determine what exact features or attributes or functions were included in the forecasted technologies. The experts had to make plausible interpretations and assumptions — usually implicitly — about the details that the forecaster intended to associate with the forecast. It is possible that different experts made different implicit assumptions, which is a limitation of the study.

Another challenge had to do with the fact a "technology" forecast can refer to at least three different categories of technological progress. In making a forecast, a forecaster can have in mind a component technology that constitutes a valuable development but cannot provide a function to a user until integrated into a full system intended for end-use. Alternatively, a forecaster may make a forecast about a fully integrated system operated by the intended end-user. Or, a forecaster may make a forecast about an ultimate capability — i.e., the utility provided to the end-user — that would be achievable by some, possibly unspecified, system.

For the purposes of this study, we and our expert assessors assumed that a forecast statement referred to a system characterized by inclusion of certain technological features or characterized by an ability to perform a certain useful function. The exact nature of the system implied by a forecaster 2 or 3 decades ago could be in doubt and required interpretations and assumptions that were not always explicit or uniform among the experts; this constituted another limitation of the study's methodology.
Yet another challenge faced by the assessors was that (considering the military nature of the technologies under consideration) they had to rely on information from open sources. Even when their professional knowledge suggested that the open sources underreported the level of a technology development or, on the contrary, reported it too optimistically, the experts were constrained to provide their assessment based on the publically available knowledge.

## Analysis and Discussion

The average accuracy for all 81 forecast statements (see the Appendix) was 0.76. This number is broadly consistent with two other known forecasts for horizons longer than 20 years. In one of these two (Parente and Anderson-Parente 2011), the average accuracy was 0.72. In another (Albright 2002), the accuracy ranged from 0.80 for fields related to computers and communications down to below 0.50 for other technology fields. On the other hand, it is far higher than 0.14 reported by Fye et al. (2013).

It might be tempting to compare the accuracy of predictions of the authors-forecasters included in this study. We decided, however, that such a comparison would be neither valuable nor fair to the forecasters for a number of reasons. For example, the intent of a work of fiction is mainly its literary and entertainment value, not the accuracy of "forecasts," even if they happened to be rather prescient (in the case of Peters [1991]). Other forecasters, e.g., O'Hanlon (2000), supplied extensive discussions of uncertainties in their forecasts, which we had to ignore in this study. Thus, we are not comparing the forecasters.

On the other hand, it is instructive to compare average accuracy values for different categories of assessments (Table 4). Broadly speaking, those categories that included a higher fraction of technologies related to acquisition, processing and distribution of information tended to fair better than others that relate to more physical effects. This resonates with the observation made in Albright (2002) that 1967 forecasts made for 2000 in computers and communication stood out as about 80% correct, while forecasts in all other fields were judged to be less than about 50% correct. Albright (2002) hypothesizes that "sustained trends of increasing capabilities and declining costs of technologies used for computers and communication applications were apparent in 1967 and enabled accurate long term forecasts."

*Table 4 Average accuracy of forecasts for different categories of forecast statements.*

| Category of forecasts | Line-of-Sight Effects | Non-Line-of-Sight Effects | Protection | Platforms | Cyber and Electronic Warfare | Sensing and Information Collection | Command and Control |
|---|---|---|---|---|---|---|---|
| Average Accuracy | .50 | .75 | .67 | .66 | .94 | .95 | .81 |

Similarly to Albright (2002), we hypothesize that forecasts related to information acquisition (sensing) and processing tend to have higher accuracy than other forecasts. To quantify this observation, we compared the group of forecasts that combine the categories of Line-of-Sight Effects, Non-Line-of-Sight Effects, Protection, and Platforms (a total of 43 statements with mean accuracy of .65) with the forecasts Cyber and Electronic Warfare, Sensing and Information Collection, and Command and Control (a total of 38 statements with mean accuracy .87).

To determine whether the difference between the two means is statistically significant, we note that the overwhelming majority of statements (71 out of 81) were assessed as 0 or 1, and therefore we can approximate our data as two proportions. Assuming that all accuracy values equal or higher than .5 represent valid forecasts, we see that the first group has 30 valid forecasts out of 43 total, and the second group has 36 valid forecasts out of 38 total. Using the "N-1" Chi-squared test as recommended by Campbell (2007), we calculated the significance level P = 0.0044. If we assume that valid forecasts are those with accuracy values strictly above 0.5, than the first group has 27 valid forecasts out of 43 total, and the second group has 32 valid forecasts out of 38 total. The significance level is still respectable P = 0.035. In other words, it is highly likely that the two groups reflect two "populations" of technological phenomena that differ in terms of being amenable to forecasts.

Another way to look at the accuracy of forecasts is to consider the fraction of all forecasts that are assessed to be at least partly correct. In our dataset, we see that out of 81 forecast statements, 66 have been assessed as correct at the 0.5 level or above. By this measure, the overall accuracy of our collection of forecasts is 81%.

In addition, one can ask whether a forecast statement refers to a technology development trend that is still seen as promising and successful at the assessment time. If so, the forecast in 1990s was leading the research managers in the right direction, even if the forecasted system is not yet developed by 2020, and remains a topic of active research and development. We inspected our list of forecast statements and identified those that (1) represent an area where significant R&D advancements have occurred and also (2) represent a direction of active R&D investment at this time. We concluded that at least 72 out of 81 statements meet the above criteria. By this measure, 89% of forecast statements are seen as correct in the sense that they forecasted valid, promising directions of R&D. To recap, we found the accuracy of forecasts to be 0.76 if measured by a simple average of all assessments; 0.81 if measured by the fraction of all statements that were assessed at the level of 0.5 or above; and 0.89 if measured by the number of statements that were assessed as successful and promising R&D directions.

To review the limitations of this research, we should first refer to several challenges of expert assessments mentioned in the previous section. The assessors had to decipher the statements of forecasters that were often brief, vague, and dependent on the context of debates in the military technology community of 2–3 decades ago. There was inevitable subjectivity in assessors' interpretations of not only the meaning of the forecast statements, but also of the degree of completeness and effectiveness of the functions that forecasters expected from the systems they forecasted. Because assessors were required to rely on information available in open literature (not always fully reliable or accurate), there were inevitable differences in assessors' familiarity with the literature and in their interpretations of maturity and readiness of the systems described in open literature. Overall, these challenges were typical of all literature in technology forecasting. It is interesting to note that the inter-rater reliability was as high as it was, in spite of these challenges.

Perhaps a far more profound limitation — also common to technology forecasting literature — is the fact that this study does not attempt a quantitative analysis of false negatives, i.e., on the failures of the forecasters to predict technologies that did in fact materialize. We see this as an important subject for further research.

## Conclusions

Military technology forecasts by their nature are primarily long term, with many useful horizons being longer than 20 years. Thus, it is important to consider the accuracy of such long-term forecasts for the purposes of military technology development and procurement, particularly in critical management decisions on allocating funds to long-term research topics.

The key finding and contribution of this study is that the accuracy of forecasts, by multiple authors, made in the 1990s with the specific horizon of 2020 is respectably high. When measured by the expert-assessed accuracy of the forecast statements, the average accuracy is 0.76. It is broadly comparable to two other known forecasts with horizons over 20 years, Parente and Anderson-Parente (2011) and Albright (2002), and it is much higher than indicated in some other works, e.g., Fye et al. (2013).

The accuracy of the forecasts under consideration in this study is even higher if we assess them by whether a forecast statement refers to a technology development trend that is seen as promising and has exhibited a significant progress by this time. We find that by this measure, 89% of forecast statements are correct in the sense that they forecasted valid, promising directions of R&D. In the 1990s, these forecasts correctly pointed to technology directions that have shown progress and continue to be active areas of development and investments.

Another significant finding is that some technology categories exhibited much higher forecast accuracy than others. Specifically, we found that there is a major difference, with strong statistical significance, in accuracies of forecasts related to "informational" (i.e., Cyber and Electronic Warfare, Sensing and Information Collection, and Command and Control) and "physical" (i.e., Line-of-Sight Effects, Non-Line-of-Sight Effects, and Protection and Platforms) technologies. The latter represented a total of 43 statements with mean accuracy of 0.65, while the former represented a total of 38 statements with mean accuracy of 0.87. This is broadly similar to, and corroborates the findings of, Albright (2002). The reasons for such significant differences in accuracies of forecasts are far from obvious; these are likely to derive from differences in the nature and maturity of technologies involved and should be a subject of further research.

Other findings and contributions of this research have to do with methodological aspects of forecast accuracy assessments. First, we consider it desirable to perform such assessments on a combination of forecasts from multiple forecasters, rather than a set of forecasts from one author. This ameliorates the concern that an exceptionally prescient forecast might skew the expectations of accuracy, as well as the impression that one must pick such an exceptional forecaster in order to obtain a sufficiently accurate forecast.

Regarding the approach to performing the assessments, we did not find it too difficult to assess, at the horizon time, whether a technology has reached a certain level of readiness. It is a significantly easier and more objective assessment than the one that deals with impact or level of adoption of the technology. The former approach focuses on technology itself; the latter unduly conflates technology maturity with market and competitive conditions. We recommend, therefore, using technology readiness as a key measure for assessing whether a forecast is successful. In addition, the "partial credit" assessments of the kind we attempted are also useful and were found to be easy to apply. It is desirable to engage multiple expert assessors (in our case up to 10) as this is likely to improve the overall quality of the assessments.

Even though the focus of the study was on military technology, constraining this study to information available in open literature has not been found to be a significant handicap. This has to do in part with the long-term nature of the forecasts — at the 20–30 year horizon — considered in the study. Forecasting at such long horizons naturally occurs primarily in open literature, even for military technologies. The assessment of accuracy — particularly the assessments of actual levels of technology readiness of systems under consideration — could be more challenging when dealing with military technology; in practice, however, it was found to be a relatively benign difficulty, except in a few cases that had to be omitted from this study.

In view that military technologies represent a large segment of the economy, and an existential factor in the life of a nation, one should welcome a greater volume of research that explores long-term forecast accuracy for military technologies. Future work should include creation of a system and a process for continuous forecasting and assessments of military technologies, perhaps somewhat similar to what is undertaken by the long-running TechCast project (Halal 2013).

Such a process for continuous assessment would be less effective if assessments by different groups of researchers could not be compared because of differences in their respective assessment methodologies. Establishing a consistent, standardized set of methodologies appropriate for uniform use within the technology forecasting community would be a highly desirable advance and a suitable topic for future work.

Another major topic for future work on the methodology of forecasting is the assessment of the implied false negatives — the technologies that are *not* predicted but do appear by the horizon time — in technology forecasts. Failure to forecast a technology (false negative) is just as serious as an incorrect forecast that a technology would appear. If a country fails to anticipate a critical military technology and fails to make appropriate investments, it may face a catastrophic surprise at the hands of an adversary who did acquire the technology. Future work on this topic would be of significant value.

The long-term future of technology is not unknowable. To the contrary, as this study suggests, the long-term forecasts of technology can be quite accurate. As such, they can be valuable contributors to informed decisions of research management.

# About the Authors

Dr. Alexander Kott serves as the U.S. Army Research Laboratory's (ARL) Chief Scientist in Adelphi, Maryland. In this role he provides leadership in the development of ARL technical strategy, maintaining technical quality of ARL research, and representing ARL to the external technical community. From 2009-2016 he was the Chief, Network Science Division, Computational and Information Sciences Directorate at ARL. Prior to that, from 2003-2008, he served as a Defense Advanced Research Programs Agency (DARPA) Program Manager.

Dr. Philip Perconti is a member of the Senior Executive Service and serves as the Director of the U.S. Army Research Laboratory (ARL), the Army's premier laboratory for basic and applied research and analysis. The Laboratory consists of approximately 2,000 civilian and military employees with an annual budget of over $1 billion. Prior to this, Dr. Perconti served as the Director of the Sensors & Electron Devices Directorate of the ARL. Earlier, he served as the Director, Science and Technology Division, US Army CERDEC Night Vision and Electronic Sensors Directorate (NVESD), and the Director, Electronics & Photonics Technology Office, National Institute of Standards & Technology (NIST).

# Appendix: Forecast Statements

| Forecasts |
|---|

Peters (P) -- forecast of 1991; Newman (N) -- forecast of 1996; Vickers (V) -- forecast of 1996; Scales (S) -- forecasts of 1997; O'Hanlon (O) -- forecast of 1999

**Line-of-Sight Effects**

Laser weapons will **not** be available on ground or air platforms (H)

High-power laser installed on a combat air platform will be capable of destroying an opponent air platform (P, N)

Electromagnetic gun will operate on an aerial gunship (P)

Electromagnetic guns will **no**t be available on ground or air platforms (H)

A tank munition will be laser-guided (H)

A gun will automatically acquire a target and automatically fire (P)

An RF weapon installed on an aircraft will cause wide-area impact on ground warfighters (P)

Infantry soldier will be able to carry a miniature precision weapon that can destroy a tank (S)

An infantryman's gun will carry a sensor that will project targets onto the heads-up display for fire against hard-to-see objects (N)

**Non-Line-of-Sight Effects**

A UAV will carry a laser designator (H)

Sub-munitions will home on a tank or other target autonomously (H)

Loitering sub-munitions will home on a tank or other target autonomously (N)

Swarms of armed UAVs (loitering munitions) will be able to destroy numerous ground targets (S, H, V)

A missile will be long range, maneuverable, and partly autonomous (H, V)

Standoff mines will be remotely triggered via a network (H)

| |
|---|
| An alternative propellant will have much lower weight and bulk than in the 1990s (S) |
| |
| A mortar round will be precision guided (H) |
| |
| A missile will be guided by an autonomous radar for a precision attack even in bad weather (H) |
| |
| A missile will be cued from UAV sensors or remotely activated by ground forces (V) |
| |
| A cruise missile will be stealthy (V) |
| |
| A system will use automated target recognition (N) |
| |
| **Protection** |
| |
| A drilling (boring) machine will produce underground facilities (H) |
| |
| A system will provide Theater Missile Defense (H) |
| |
| Protection from chemical and biological agents will be incorporated into the uniform (N) |
| |
| **Platforms** |
| |
| UAVs will be used for a range of functions, including weapon delivery and EW (V, H) |
| |
| UAVs will roam at high altitudes for days (N) |
| |
| A tank (or gun, or mortar) will operate unmanned (H) |
| |
| A robotic (unmanned?) ground vehicle will be used as a scout (P, H) |
| |
| UAVs will be used as scouts (P) |
| |
| Mine-resistant vehicles will be widely used (H) |
| |
| A combat vehicle will have an alternative power source, be fuel efficient, and ultra-reliable (S) |
| |
| Battery-driven battlefield vehicles will not be available (H) |

A hybrid power generation system and energy storage will have much lower weight and bulk than in the 1990s (S)

A vehicle will be powered by hydrogen-fuel cell (H)

A ground craft will be composed of advanced lightweight materials, with significantly higher firepower, mobility, and speed than in the 1990s (S)

Tanks will be no lighter than in the 1990s (H)

A main battle tank will include an unmanned gun mount; 3-person crew sitting in a compartment in forward hull, with observations via sensors (P)

A variety of platforms, including helicopters, will be stealthy (V, H)

A combat air platform will use tilt-rotor (P, S)

A combat vehicle will maneuver within folds of earth without contact with the ground (S)

A light ground combat vehicle with crew inside will maneuver by being lifted by an airframe and transported at speeds up to 200 km/hr at distances over 1500 km (S)

Stealthy parafoils will be used for precision resupply (V)

Exoskeleton will serve as an armored personal mobility system (V)

## Cyber and Electronic Warfare

Adaptable, time-phased computer viruses will be able to access and destroy remote computer nodes via a computer network (P, V)

An RF weapon will be able to attack communication networks (H)

An electromagnetic pulse (EMP) weapon will provide area effects (V)

A weapon will be based on delivering a high-power microwave (HPM) beam (V)

A cyberattack system will be able to deceive an enemy information system regarding the force locations or posture (V)

| |
|---|
| Cyber and EW systems will be able to induce information overload on enemy C2 (V) |
| A UAV will function as a jammer (P) |
| A high-power jamming system will be able to destroy enemy electronics (P) |
| **Sensing and Information Collection** |
| Radars will be ground- and wall-penetrating (H) |
| A foliage-penetrating radar will be able to see under jungle canopy (H) |
| Multispectral sensors will be available (V) |
| Cameras will be able to see in bad weather, day or night (N) |
| Miniature UAVs and unmanned ground sensors will be able to track the enemy in urban and other complex environments (H) |
| Air-delivered unmanned ground sensors will detect and report enemy forces by sensing heat, movement, sound, and fumes (N) |
| Nets of ground sensors will connect to constellation of UAVs at various altitudes and will serve as unblinking eye (S) |
| Multiple UAVs will form a "web" over the battlefield, relaying pictures of the area (N) |
| Satellites and platforms like JSTARS will track enemy forces (H) |
| An unmanned sensor will process data on board and recognize targets (H) |
| A sniper detector will perform detection after the first shot (H) |
| Front-line troops will be able to use palm-sized UAVs to relay pictures from several hundred yards forward (N) |
| Soldier's helmet will include a camera, and its live video will be fed into the network (N) |
| Soldiers will carry a combat identification device to pre-empt fire by friendly forces (N) |

## Command and Control

Computers will be widely used on the battlefield (H)

Essential battlefield data will be collected into a digital warehouse (S)

Information about nearly everything from satellites to UAVs to UGSs will be networked and integrated over the entire theater into a real-time picture (N)

Computers will be able to process massive warehouses of information, such as millions of pictures, filter and assemble all into one mosaic (N)

The battlefield picture will receive frequent automatic updates, on the order of every 20 seconds (N)

A persistent observation systems will hover over a fighting force, protecting it from tactical surprise and calling for fires within seconds (S)

A network of sensors and shooters will be able to destroy the enemy beyond the range of the enemy's tactical weapons (S)

Dissimilar networks will connect over wide areas and down to the individual warrior (V)

Target information will be rapidly disseminated between platforms so that a weapon itself does not need to acquire a target (H)

Networks will **not** have the capacity to distribute imagery on the battlefield (H)

UAVs will provide backup to satellites in order to ensure networked comms (H, S)

A commander will be able to see every enemy platform on a screen, and a click on an icon will provide details like UAV footage (N)

Commanders will be able to rapidly direct attacks by a click on target icons "nominated" by software (N)

Commanders will stay linked to the common picture even when they're on the move (N)

| |
|---|
| In a gunship, extensive information and sensor data will be fused on displays (P) |
| |
| Soldier's helmet will have a heads-up display with enemy targets and friendlies, obtained from various sources (N,H) |
| |
| **COUNT: 81** |